%

\input epsf
\magnification=1200
\overfullrule=0mm
\centerline{\bf Matrices coupled in a chain. I. Eigenvalue correlations}
\vskip 3mm

\centerline{Bertrand Eynard\footnote\S {e-mail address: 
eynard@spht.saclay.cea.fr } and 
Madan Lal Mehta\footnote* {Member of 
Centre National de la Recherche Scientifique, France}
\footnote\dag{e-mail address: mehta@spht.saclay.cea.fr}}
\centerline{CEA/Saclay, Service de Physique Th\'eorique}
\centerline{F-91191 Gif-sur-Yvette Cedex, FRANCE}
\vskip 1cm
\noindent {\bf Abstract.} The general correlation function for the 
eigenvalues of $p$ complex hermitian $n\times n$ matrices coupled in 
a chain is 
given as a single determinant. For this we use a slight 
generalization of a theorem of Dyson.
\vskip 19mm

\noindent {\bf 1. Introduction}
\vskip 3mm

\noindent The probability density ${\rm exp}[-{\rm tr}\ V(A)]$ for the 
elements of a $n\times n$ matrix $A$ is known to give rise to the 
probability density [1]
$$ F(x_1,\dots,x_n) \,\propto\, \exp\left[-\sum_{i=1}^n V(x_i)\right] 
\prod_{1\le i<j\le n} \vert x_i-x_j \vert^\beta \eqno (1.1) $$
for its eigenvalues $x_1$, ..., $x_n$. Here $V(x)$ is a real polynomial 
of even 
order, the coefficient of the highest power being positive and $\beta$ is 
the number of real components of a 
general element of $A$, i.e. $\beta=1$ if $A$ is real symmetric, $\beta=2$ 
if $A$ is complex hermitian and $\beta=4$ if $A$ is quaternion self-dual.

The case of coupled matrices may be represented by a graph where each matrix 
is represented by a point, and two points representing matrices $A$ and $B$ 
are joined by a line if the coupling factor $\exp[c\ {\rm tr}(AB)]$ is 
present in the probability density. When several matrices are 
coupled the probability density for the eigenvalues is known only 
in the case where these matrices are complex hermitian and the graph has 
a tree structure, i.e. does not have a closed path.

In what follows we will consider the simplest case of a tree, i.e. that of 
a chain of $p$ complex hermitian 
$n\times n$ matrices with the probability density for their elements
$$ \eqalignno{ F(A_1,\cdots , A_p)\, & \propto\, 
\exp\left[-{\rm tr}\left\{ {1\over 2}V_1(A_1)+V_2(A_2)+\cdots+V_{p-1}
(A_{p-1})+{1\over 2} V_p(A_p)\right\}\right] & \cr 
& \times\,\exp\left[ {\rm tr} \left\{ c_1A_1A_2+c_2A_2A_3+\cdots
+c_{p-1}A_{p-1}A_p \right\}\right], & (1.2) \cr}$$
where $V_j(x)$ are real polynomials of even order with positive coefficients 
of their highest powers and $c_j$ are real constants. For each $j$ the 
eigenvalues of the matrix $A_j$ are real and will be denoted 
by ${\bf x_j} := \left\{ x_{j1}, x_{j2}, ..., x_{jn}\right\}$. 
The probability density for 
the eigenvalues of all the $p$ matrices resulting from Eq. (1.2) is [2-5]
$$ \eqalignno{ & F({\bf x_1}; ...;{\bf x_p}) & \cr
& = C \exp\left[-\sum_{r=1}^n\left\{ {1\over 2}V_1(x_{1r})+
V_2(x_{2r})+\cdots+V_{p-1}(x_{p-1r})
+{1\over 2} V_p(x_{pr})\right\} \right] & \cr
& \times \prod_{1\le r<s\le n}(x_{1r}-x_{1s})(x_{pr}-x_{ps}) & \cr
& \times \det\left[e^{c_1x_{1r}x_{2s}}\right] 
 \det\left[e^{c_2x_{2r}x_{3s}}\right]\cdots\det\left[e^{c_{p-1}
x_{p-1r}x_{ps}}\right] & (1.3) \cr
& = C\left[ \prod_{1\le r<s\le n}(x_{1r}-x_{1s})(x_{pr}-x_{ps}) \right]
\left[ \prod_{k=1}^{p-1} \det\left[w_k(x_{kr},x_{k+1s})\right]_{r,s=1,...n}
\right], & (1.4) \cr}$$
where
$$w_k(\xi,\eta):=\exp\left[-{1\over 2}V_k(\xi)-{1\over 2}V_{k+1}(\eta)
+c_k\xi\eta \right], \eqno (1.5)$$
and $C$ is a normalisation constant such that the integral of $F$ over the 
$np$ variables $x_{ir}$ is one.
We will be interested in the correlation functions 
$$ \eqalignno{ & R_{k_1,...,k_p}(x_{11},...,x_{1k_1};...;x_{p1},...,x_{pk_p})
& \cr
&  := \int  F({\bf x_1};...;{\bf x_p}) 
\prod_{j=1}^p {n!\over (n-k_j)!}
\prod_{r_j=k_j+1}^n dx_{jr_j}. & (1.6) \cr} $$
This is the density of ordered sets of $k_j$ eigenvalues of $A_j$ within 
small intervals around $x_{j1}$, ..., $x_{jk_j}$ for $j=1$, 2, ..., $p$.
Here and in what follows, all the integrals 
are taken over $-\infty$ to $\infty$.

The case $p=2$ was considered earlier [6,10]; the expressions given in~[6], 
though correct, can be put in a much simpler form:
the general answer can be written as a single $m\times m$ 
determinant with $m=k_1+k_2+\cdots+k_p$. 
The result is given in section 2 and the proof in section 3. 
Our result is a generalization of Dyson's for a single hermitian matrix 
[7], according to which the correlation function of $k$ eigenvalues is 
given by a $k\times k$ determinant:
$$ R_k(x_1,\dots,x_k)=\det \left[ K (x_i,x_j) \right]_{i,j=1,..., k }. 
\eqno(1.7)$$
\vskip 8mm

\noindent {\bf 2. General correlation function}
\vskip 3mm

\noindent To express our result we need some notations.
Recall that a polynomial is called monic when the coefficient of the 
highest power is one.
With a monic polynomial $P_j(\xi)$ of degree $j$ let us write
$$ P_{1j}(\xi) := P_j(\xi), \eqno(2.1)$$
and recursively,
$$P_{ij}(\xi):=\int P_{i-1j}(\eta)w_{i-1}(\eta,\xi)d\eta,\ \ \ 2\le i\le p. 
\eqno(2.2)$$
Similarly, with a monic polynomial $Q_j(\xi)$ of degree $j$ we will write 
$$ \eqalignno{ Q_{pj}(\xi) & := Q_j(\xi), & (2.3) \cr
Q_{ij}(\xi) & := \int w_i(\xi,\eta)Q_{i+1j}(\eta) d\eta, 
\ \ \ 1\le i\le p-1.  & (2.4) \cr}$$
With arbitrary monic polynomials $P_j(\xi)$ and $Q_j(\xi)$ of 
degree $j$, $j=0$, 1, 2, ..., 
we can write the product of differences as $n\times n$ determinants
$$ \prod_{1\le i<j\le n} (\xi_j-\xi_i) = \det[\xi_i^{j-1}] = 
\det[P_{j-1}(\xi_i)] = \det[Q_{j-1}(\xi_i)]. \eqno(2.5) $$
Now if the polynomials $V_j$ and the constants $c_j$ are such that the 
moment matrix $[M_{ij}]$, $i,j=0$, 1, ..., $n$ is positive definite for
every $n$, where
$$ M_{ij}:=\int \xi^i\,(w_1*w_2*...*w_{p-1})(\xi,\eta)\,\eta^j\, d\xi d\eta, 
\eqno (2.6)$$
and
$$ (w_{i_1}*w_{i_2}*\cdots*w_{i_k})(\xi,\eta) 
:= \int w_{i_1}(\xi,\xi_1)w_{i_2}(\xi_1,\xi_2)... w_{i_k}(\xi_{k-1},\eta) 
d\xi_1... d\xi_{k-1}, \eqno (2.7) $$
then it is always possible to 
choose the polynomials $P_j(\xi)$ and $Q_j(\xi)$ such that
$$\int P_j(\xi)(w_1*w_2*... *w_{p-1})(\xi,\eta)
Q_k(\eta) d\xi d\eta = h_j \delta_{jk}, \eqno(2.8)$$
i.e. they are orthogonal with a non-local weight. This 
means that the functions $P_{ij}(\xi)$ and $Q_{ij}(\xi)$, which are not 
necessarily polynomials, are orthogonal 
$$\int P_{ij}(\xi) Q_{ik}(\xi)d\xi = h_j \delta_{jk}, 
\eqno(2.9)$$
for $i=1$, 2, ..., $p$ and $j,k=0$, 1, 2, ... .

\noindent Now define
$$ K_{ij}(\xi,\eta) := H_{ij}(\xi,\eta)-E_{ij}(\xi,\eta), \eqno(2.10)$$
where
$$ \eqalignno{ H_{ij}(\xi,\eta) & := \sum_{\ell=0}^{n-1}
{1\over h_\ell} Q_{i\ell}(\xi) P_{j\ell}(\eta), & (2.11) \cr
E_{ij}(\xi,\eta) & := \left\{ \matrix {0,\hfill & {\rm if}\ i\ge j, \hfill \cr
w_i(\xi,\eta), \hfill & {\rm if}\ i= j+1, \cr
(w_i*w_{i+1}*\cdots*w_{j-1})(\xi,\eta), & {\rm if}\ i<j+1. \cr} \right. 
 & \matrix{ (2.12a) \cr (2.12b) \cr (2.12c) \cr} \cr}$$

\noindent {\bf Theorem.} The correlation function (1.6) is equal to 
$$ \eqalignno{ & R_{k_1,...,k_p}(x_{11},...,x_{1k_1};...;x_{p1},...,x_{pk_p}) 
& \cr
& = \det \left[K_{ij}\left(x_{ir}, x_{js}\right)\right]_
{i,j=1,...,p; r=1,...,k_i; s=1,...,k_j}. & (2.13) \cr}$$
This determinant has $m=k_1+\cdots+k_p$ rows and $m$ columns; the first 
$k_1$ rows and $k_1$ columns are labeled by the pair 
of indices $1r$, $r=1$, ..., $k_1$; the next $k_2$ rows and $k_2$ 
columns are labled by the pair of indices $2r$, $r=1$, ..., 
$k_2$ and so on. Each variable $x_{ir}$ appears in exactly one row 
and one column, this row and column crossing at the main diagonal. 
If all the eigenvalues of a matrix $A_j$ are not observed (are integrated   
out), then no row or column corresponding to them appears in Eq. (2.13). 
If all the $k_j$ are zero, then the correlation function is 1.
\vskip 1cm

\noindent {\bf 3. Proofs}
\vskip 3mm
 
\noindent The theorem is a consequence of the following two lemmas. 
\vskip 3mm

\noindent {\bf Lemma 1.} The $np\times np$ determinant 
$\det[K_{ij}(x_{ir},x_{js})]$, $i,j=1,...,p$; $r,s=1,...,n$, is, apart 
from a constant, equal to the probability density 
$F({\bf x_1};...;{\bf x_p})$, Eq. (1.4),
$$ \det[K_{ij}(x_{ir},x_{js})]_{\matrix{i,j=1,...,p \cr 
r,s=1,...,n\cr}}=\left(\prod_{\ell=0}^{n-1}h_\ell^{-1}\right)
C^{-1}F({\bf x_1};...;{\bf x_p}). \eqno(3.1)$$

\noindent {\bf Lemma 2.} Using the obvious notation of Eq. (2.7), 
$$ (f*g)(\xi,\eta):=\int f(\xi,\zeta)g(\zeta,\eta)d\zeta, \eqno(3.2)$$
let the $p^2$ functions $K_{ij}(x,y)$, $i,j=1,...,p$, be such that 
$$ \eqalignno{ K_{ij}*K_{jk} & = 
\left\{ \matrix{ K_{ik}, & {\rm if}\ i\ge j\ge k, \hfill \cr
 -K_{ik}, & {\rm if}\ i<j<k, \hfill \cr
 0, & {\rm otherwise}. \hfill }\right. & (3.3) \cr} $$  
Then the integral of the $m\times m$ determinant 
$\det\left[K_{ij}\left(x_{ir}, x_{js}\right)\right]$, 
($i,j=1,...,p$; $r=1,...,k_i$; $s=1,...,k_j$; $k_1\ge 0$, ...,  
$k_p\ge 0$; $m=k_1+k_2+...+k_p$), over $x_{\ell t}$ is 
proportional to the $(m-1)\times (m-1)$ determinant obtained from 
it by removing the row and the column containing the variable 
$x_{\ell t}$. The constant of proportionality is $\alpha_\ell -k_\ell +1$, 
with 
$$ \alpha_\ell =\int K_{\ell \ell }(x,x)dx. \eqno (3.4)$$
\vskip 3mm

Let us recall here a result of Dyson [7,8].

\noindent {\bf Dyson's theorem.} Let the function $K(x,y)$ be such that 
$K*K=K$, then
$$ \int \det[K(x_i,x_j)]_{i,j=1,...,n}dx_n=(\alpha-n+1)
\det[K(x_i,x_j)]_{i,j=1,...,n-1}, \eqno (3.5)$$
with
$$ \alpha=\int K(x,x)dx. \eqno (3.6)$$
Our lemma 2 above is a generalization of this one when the 
matrix elements $K(x_i,x_j)$ are replaced by $k_i\times k_j$
matrices $K_{ij}(x_{ir},x_{js})$.
\vskip 3mm                    

\noindent {\bf Proof of lemma 1.} Consider the $np\times np$ matrix 
$\left[ H_{ij}\left(x_{ir},x_{js}\right)\right]$, Eq. (2.11), 
the rows of which are denoted by the pair of indices $ir$ and the 
columns by $js$; $i,j=1$, 
..., $p$; $r,s=1$, ..., $n$. This matrix can be written as the product 
of two rectangular matrices $\left[Q_{i\ell}(x_{ir})\right]$ and 
$\left[P_{j\ell}(x_{js})/h_\ell\right]$ respectively of sizes $np\times n$ 
and $n\times np$, with $\ell=0$, 1, ..., $n-1$. The rows of the first 
matrix $[Q_{i\ell}(x_{ir})]$ are numbered by the pair $ir$ and its 
columns by $\ell$. For $[P_{j\ell}(x_{js})/h_\ell]$ the rows are numbered 
by $\ell$ and the columns by $js$. 
Cutting the matrix $\left[H_{ij}(x_{ir},x_{js})\right]$ into $n\times n$ 
blocks, we can write 
$$ \eqalignno{ H & = \left[
\matrix{\bar Q_1\bar P_1 & \bar Q_1\bar P_2 & \cdots & \bar Q_1\bar P_p \cr
        \bar Q_2\bar P_1 & \bar Q_2\bar P_2 & \cdots & \bar Q_2\bar P_p \cr
        \cdots    & \cdots    & \cdots & \cdots    \cr
        \bar Q_p\bar P_1 & \bar Q_p\bar P_2 & \cdots & \bar Q_p\bar P_p \cr } 
\right]_{np\times np} & \cr
& = \left[ \matrix{ \bar Q_1 \cr \bar Q_2 \cr \vdots \cr 
\bar Q_p \cr}\right]_{np\times n} 
 \left[ \matrix{ \bar P_1 & \bar P_2 & \cdots & \bar P_p \cr}
\right]_{n\times np}, & (3.7) \cr}$$
where 
$\left[\bar Q_i\right]_{r\ell} := \left[ Q_{i\ell}(x_{ir})\right]$ and  
$\left[ \bar P_{j}\right]_{\ell s} := \left[P_{j\ell}(x_{js})/h_\ell\right]$ 
are $n\times n$ matrices. 
The rows of $\bar P_1$ and the columns of $\bar Q_p$ contain distinct monic 
polynomials, their ranks are therefore $n$. Thus the rank of 
$\left[H_{ij}(x_{ir},x_{js})\right]$ is $n$. As seen from Eq. (3.7) 
above, the first $n$ columns of this 
matrix are linearly independent, while the remaining $n(p-1)$ columns 
can be linearly expressed in terms of the first $n$ columns. 
In view of Eqs. (2.10) and (2.12a) the first $n$ columns of 
$\left[H_{ij}(x_{ir},x_{js})\right]$ are identical 
with the first $n$ columns of $\left[K_{ij}(x_{ir},x_{js})\right]$. The 
determinant of the later 
is therefore not changed if we subtract from its last $n(p-1)$ columns the
corresponding $n(p-1)$ columns of the former. Thus
$$ \eqalignno{
\det\left[K_{ij}(x_{ir},x_{js})\right] & = \det \left[
\matrix{H_{i1}(x_{ir},x_{1s}) & -E_{ij}(x_{ir},x_{js}) \cr}\right]
_{\matrix{i=1,...,p; \ j=2,...,p; \cr r,s=1,2,...,n; \hfill \cr}} & \cr
 & =\det \left[
\matrix{ H_{11} & -E_{12} & -E_{13} & \cdots & -E_{1p} \cr
H_{21} & 0 & -E_{23} & \cdots & -E_{2p} \cr 
\vdots &  \vdots & \vdots & \ddots & \vdots \cr
H_{p-1\, 1} & 0 & 0 & \cdots & -E_{p-1\, p} \cr
H_{p1} & 0 & 0 & \cdots  & 0 \cr }\right] . & (3.8)\cr}$$
From Eq. (2.12a) the last $n$ rows of this matrix corresponding 
to $i=p$ have non-zero elements only in 
the first $n$ columns; also the matrix $\left[ E_{ij}(x_{ir},x_{js})\right]$ 
is block triangular, $E_{ij}(\xi,\eta)$ being zero for $i\ge j$. 
Therefore,
$$ \eqalignno{ \det \left[ K_{ij}(x_{ir}, x_{js})\right] 
& = \det \left[H_{p1}(x_{pr},x_{1s})\right]\det\left[
E_{ij}(x_{ir},x_{js})\right]_{\matrix{i=1,...,p-1;\ j=2,...,p; \cr
r,s=1,2,...,n; \hfill \cr}} & \cr
& = \det \left[H_{p1}(x_{pr},x_{1s})\right]
\prod_{j=2}^p\det\left[
E_{j-1\, j}(x_{j-1r},x_{js})\right] & \cr
& = \left(\prod_{\ell=0}^{n-1}h_\ell^{-1}\right) \det\left[
Q_{p\ell}(x_{pr})\right]\det\left[P_{1\ell}(x_{1s})\right]
\prod_{j=1}^{p-1}\det\left[ w_j(x_{jr},x_{j+1s})\right], & \cr
& & (3.9) \cr}$$
and from Eqs. (1.4) and (2.5) one gets Eq. (3.1). This ends the proof. 

The learned reader will have recognized that the above $E_{ij}$'s 
play the same role as the $\varepsilon$ in Dyson's proof in the case 
of a single matrix of the circular orthogonal ensemble [7,9].
\vskip 3mm

\noindent {\bf Proof of Lemma 2.} 
We want to integrate the $m\times m$ determinant 
$\left[K_{ij}(x_{ir},x_{js})\right]$ over $x_{\ell t}$. We can write 
the expansion of the determinant as a sum over $m!$ permutations, writing 
these permutations as a product of mutually exclusive cycles. The variable 
$x_{\ell t}$ occurs in the row and the column labeled by $\ell t$ 
(recall that rows and columns are labeled by a pair of indices). 
If $\ell t$ forms a cycle by itself, 
then by Eq. (3.4) integration over $x_{\ell t}$ gives a factor 
$\alpha_\ell $, 
and its coefficient is just the expansion of the $(m-1)\times (m-1)$ 
determinant obtained by removing the row and the column containing 
$x_{\ell t}$. If $\ell t$ occurs in a longer cycle, say in the 
permutation $\sigma=(ir, \ell t, js,\cdots)(\cdots)\cdots$, 
then from Eq. (3.3) integration over $x_{\ell t}$ decreases the length of 
the cycle containing $\ell t$ by one, giving the permutation 
$\sigma'=(ir, js,\cdots)(\cdots)\cdots$, and multiplies it by a factor 
$+1$ if $i\ge \ell \ge j$, by $-1$ if $i<\ell <j$ and by $0$ otherwise. 
Also, the cycle lengths differing by one, the permutation $\sigma$ has 
a sign opposite to that of $\sigma'$. 
So the question is, given the permutation $\sigma'$, in how many ways 
can one insert $\ell t$ in any of its cycles with the algebraic 
weights $+1$, $-1$ and $0$ to get the permutation $\sigma$. 

Let us represent the cycles of permutations by a graph. Only 
the first indices $i$, $j$,~...  of each pair of indices will 
be plotted against the place number where they occur.
For example, the cycle $(42,26,36,15,24,22)$ is represented on 
figure 1, where points at successive heights $ 4,2,3,1,2,2 $ are 
joined successively by line segments or ``sides".
Note that we identify the $7^{\rm th}$ point and the first one.
Permutation $\sigma'$ is thus represented by a certain number of 
closed directed polygons corresponding to its mutually exclusive cycles.
Addition of $\ell t$ in one of the cycles of $\sigma'$ amounts to the 
addition of a point at a height $\ell $ in the corresponding polygon.
If this added point lies on a non-ascending side, then the coefficient 
multiplying the corresponding $\sigma$ is $+1$, if it lies on an  
ascending side, the coefficient is $-1$, and this coefficient is 
zero otherwise.
In other words, each downward crossing of the line at height $\ell $, with 
or without stops, contributes a factor $+1$, each point on this height 
contributes a factor $+1$ and each upward crossing, with or without 
stops, contributes a factor $-1$. 
The graph of $\sigma'$ consisting of closed loops, the number of upward 
crossings is equal to the number of downward crossings at any height.

\vskip 1cm
\hbox{ \vbox{ \hbox{\hskip3cm {\epsfysize=5cm \epsfbox{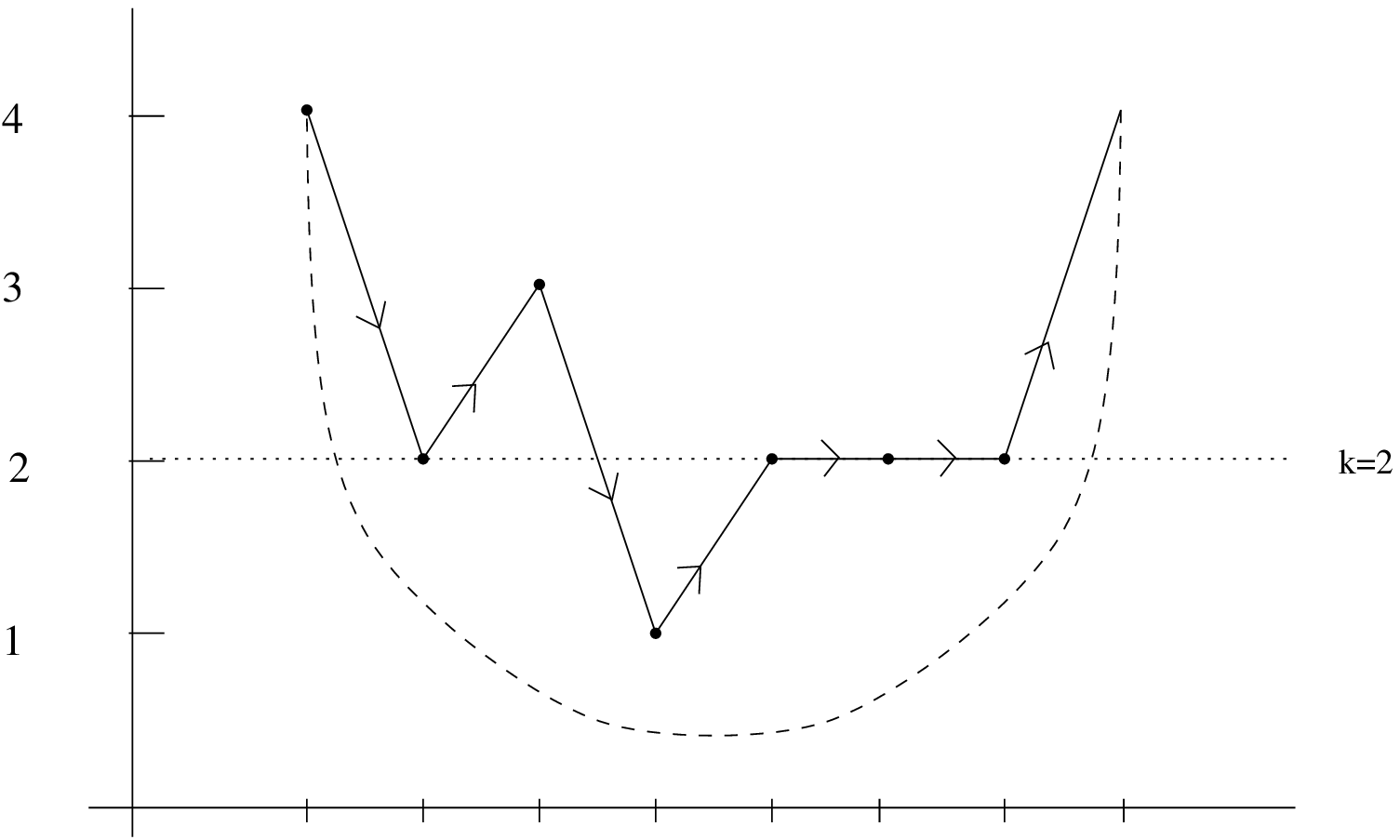}}} 
{\hskip3cm Figure 1: the permutation $\sigma'=(4,2,3,1,2,2)$.} } }
\vskip1cm
 
\noindent The algebraic sum of all such coefficients is thus seen to 
be the number 
of points at height $\ell $ in the graph of $\sigma'$, i.e. it is $k_\ell-1$.
Also the permutations $\sigma$ and $\sigma'$ have opposite signs, since 
only one of their cycle lengths differs by unity. Thus
$$\int \det\left[K_{ij}(x_{ir},x_{js})\right]_{m\times m} dx_{\ell t} 
= (\alpha_\ell-k_\ell+1) \det\left[K_{ij}(x_{ir},x_{js})\right]_{(m-1)\times 
(m-1)}, \eqno(3.10)$$ 
where the integrand on the left hand side is a $m\times m$ determinant, 
$i,j=1$, ..., $p$; $r=1$, ..., $k_i$; $s=1$, ..., $k_j$; 
$m=k_1+k_2+\cdots+k_p$ and the 
result on the right hand side is the $(m-1)\times (m-1)$ determinant obtained 
from the integrand by removing the row and the column containing the 
variable $x_{\ell t}$. This ends the proof.

Using Eqs. (2.9-2.12), one verifies that the $H_{ij}$ and 
$E_{ij}$ satisfy the following relations
$$ \eqalignno{ H_{ij}*H_{jk} & = H_{ik}, & (3.11) \cr
H_{ij}*E_{jk} & = \left\{ \matrix{ H_{ik}, & {\rm if}\ j<k, \cr
 0, & {\rm if}\ j\ge k, \cr} \right.  & (3.12) \cr
E_{ij}*H_{jk} & = \left\{ \matrix{ H_{ik}, & {\rm if}\ i<j, \cr
 0, & {\rm if}\ i\ge j, \cr} \right. & (3.13) \cr
E_{ij}*E_{jk} & = \left\{ \matrix{ E_{ik}, & {\rm if}\ i<j<k, \hfill \cr
 0, & {\rm if\ either}\ i\ge j\ {\rm or}\ j\ge k. \cr}\right. & (3.14) \cr}$$
This implies for the $K_{ij}$ the relations (3.3). Also 
$$ \alpha_\ell = \int K_{\ell \ell }(\xi,\xi) d\xi = n. \eqno(3.15) $$
Using lemma 1 once and lemma 2 several times one gets the normalization 
constant $C$, 
$$C= (n!)^{-p}\prod_{\ell=0}^{n-1}h_\ell^{-1}. \eqno(3.16)$$
Again with repeated use of lemma 2 and from the definition, Eq. (1.6), of 
the correlation function 
$R_{k_1,...,k_p}(x_{11},...,x_{1k_1}; ...;x_{p1},...,x_{pk_p})$, 
one gets Eq. (2.13). 
\vskip 8mm

\noindent {\bf 4. Conclusion}
\vskip 3mm

\noindent The correlation functions of eigenvalues of a chain of random 
hermitian matrices can thus be written in a very compact form as a 
single determinant.
This result may be used to study the large $n$ limit of correlations 
between eigenvalues [10,11].

\vskip 1cm

\noindent {\bf Acknowledgements} 
\vskip 3mm

\noindent We are thankful to J.-M. Normand and G. Mahoux 
for some discussions and specially to the former for reading and criticising  
the manuscript which helped us to make the presentation more readable and 
reduce the number of errors.
\vskip 1cm

\noindent {\bf References}
\vskip 3mm

\item{1.} See for example, M.L. Mehta, {\it Random Matrices}, Academic Press,
San Diego, CA, U.S.A., 1991, Chapter 3. Here only the case $V(x)=x^2$ is 
considered, but the same method applies when $V(x)$ is any real polynomial of 
even order. 

\item{2.} See for example, reference 1, appendix A.5.

\item{3.} C. Itzykson and J.-B. Zuber, The planar approximation II, J. Math. 
Phys. 21 (1980) 411-421. 

\item{4.} M.L. Mehta, A method of integration 
over matrix variables, Comm. Math. Phys. 79 (1981) 327-340.

\item{5.} M.L. Mehta, {\it Matrix Theory}, Les Editions de Physique, 
Les Ulis, France, 1989, chapter 13.2

\item{6.} M.L. Mehta and P. Shukla, Two coupled matrices: eigenvalue 
correlations and spacing functions, J. Phys. A: Math Gen. 27 (1994) 
7793-7803.

\item{7.} F.J. Dyson, Correlations between the eigenvalues of a random 
matrix, Comm. Math. Phys. 19 (1970) 235-250.

\item{8.} See for example reference 1, theorem 5.2.1, or reference 5, 
theorem 8.10.

\item{9.} See for example reference 1, section 6.3, or section 10.3.

\item{10.} B. Eynard, Eigenvalue distribution of large random matrices, 
from one matrix to several coupled matrices, Nuc. Phys. B (1997), (to 
appear), xxx, cond-mat/9707005.

\item{11.} B. Eynard, Correlation functions of eigenvalues of 
multi-matrix models, and the limit of a time depedent matrix, In preparation.

\end